\documentclass[prl,twocolumn,amsmath,amssymb,superscriptaddress, showpacs]{revtex4-1}

\usepackage{amsmath} 
\usepackage{graphicx} 

\usepackage{bbold}
\usepackage{hyperref} 
\usepackage{tabularx}
\usepackage{bibunits}
\usepackage[capitalise]{cleveref}

\defaultbibliography{QuantumCryoNet}

\crefname{equation}{Eq.}{Eqs.}
\Crefname{equation}{Equation}{Equations}
\crefname{figure}{Fig.}{Figs.}
\Crefname{figure}{Figure}{Figures}
\crefname{section}{Sec.}{Secs.}
\Crefname{section}{Section}{Sections}

\newcommand{\ket}[1]{\left| #1 \right>}
\newcommand{\bra}[1]{\left< #1 \right|}

\newcommand{\av}[1]{\langle #1 \rangle}

\DeclareMathOperator{\sech}{sech}
\DeclareMathOperator{\Tr}{Tr}

\begin{document}

\title{
Microwave Quantum Link between Superconducting Circuits\\ 
Housed in Spatially Separated Cryogenic Systems}

\author{P.~Magnard}
\email{paul.magnard@phys.ethz.ch}
\affiliation{Department of Physics, ETH Z\"urich, CH-8093 Z\"urich, Switzerland}

\author{S.~Storz}
\affiliation{Department of Physics, ETH Z\"urich, CH-8093 Z\"urich, Switzerland}

\author{P.~Kurpiers}
\affiliation{Department of Physics, ETH Z\"urich, CH-8093 Z\"urich, Switzerland}

\author{J.~Sch\"ar}
\affiliation{Department of Physics, ETH Z\"urich, CH-8093 Z\"urich, Switzerland}

\author{F.~Marxer}
\affiliation{Department of Physics, ETH Z\"urich, CH-8093 Z\"urich, Switzerland}

\author{J.~L\"utolf}
\affiliation{Department of Physics, ETH Z\"urich, CH-8093 Z\"urich, Switzerland}

\author{J.-C.~Besse}
\affiliation{Department of Physics, ETH Z\"urich, CH-8093 Z\"urich, Switzerland}

\author{M.~Gabureac}
\affiliation{Department of Physics, ETH Z\"urich, CH-8093 Z\"urich, Switzerland}

\author{K.~Reuer}
\affiliation{Department of Physics, ETH Z\"urich, CH-8093 Z\"urich, Switzerland}

\author{A.~Akin}
\affiliation{Department of Physics, ETH Z\"urich, CH-8093 Z\"urich, Switzerland}

\author{B.~Royer}
\altaffiliation[Current address: ]{Department of Physics, Yale University, New Haven, Connecticut 06520, USA}
\affiliation{Institut Quantique and D\'epartement de Physique, Universit\'e de Sherbrooke,
Sherbrooke, Qu\'ebec J1K 2R1, Canada}

\author{A.~Blais}
\affiliation{Institut Quantique and D\'epartement de Physique, Universit\'e de Sherbrooke,
Sherbrooke, Qu\'ebec J1K 2R1, Canada}
\affiliation{Canadian Institute for Advanced Research, Toronto, ON, Canada}

\author{A.~Wallraff}
\email{andreas.wallraff@phys.ethz.ch}
\affiliation{Department of Physics, ETH Z\"urich, CH-8093 Z\"urich, Switzerland}
\affiliation{Quantum Center, ETH Z\"urich, 8093 Z\"urich, Switzerland}

\date{\today}

\begin{abstract}
Superconducting circuits are a strong contender for realizing quantum computing systems, and are also successfully used to study quantum optics and hybrid quantum systems. 
However, their cryogenic operation temperatures and the current lack of coherence-preserving microwave-to-optical conversion solutions have hindered the realization of superconducting quantum networks either spanning different cryogenics systems or larger distances.
Here, we report the successful operation of a cryogenic waveguide coherently linking transmon qubits located in two dilution refrigerators separated by a physical distance of five meters.  We transfer qubit states and generate entanglement on-demand with average transfer and target state fidelities of $85.8\,\%$ and $79.5\,\%$, respectively, between the two nodes of this elementary network. Cryogenic microwave links do provide an opportunity to scale up systems for quantum computing and create local area quantum communication networks over length scales of at least tens of meters.
\end{abstract}

\maketitle

\begin{bibunit}[apsrev4-1]
Superconducting circuits are an appealing platform to execute quantum information processing  algorithms on noisy-intermediate-scale or error-correctable quantum hardware~\cite{Preskill2018,Arute2019,Kandala2019,Ofek2016,Andersen2020b}, and, also, to study fundamental quantum phenomena~\cite{VanLoo2013,Hacohen-Gourgy2016,Cottet2017,Minev2019}.
Today's state-of-the-art superconducting quantum processors contain a few dozen qubits on a single chip, held at cryogenic temperatures in individual dilution refrigerators. Efforts in qubit integration and packaging~\cite{Bejanin2016,Das2018a,Foxen2018,Lei2020} will
likely extend the scale of these processors to thousands of qubits in the foreseeable future. However, limitations such as available wafer size, refrigerated space and cooling power may arise beyond that scale~\cite{Krinner2019}. Therefore, major innovations in both device integration and cryogenics are required to realize error-corected quantum computers able to tackle interesting problems intractable on high-performance computing (HPC) systems, likely requiring millions of qubits~\cite{Reiher2017,Babbush2018a}.
Networking quantum processors housed in different cryogenic nodes may provide a modular solution to scale up quantum computers beyond these limitations~\cite{Nickerson2014,Brecht2016}. 
The capabilities of quantum computers may be extended by forming clusters of networked processors housed in individual cryogenic modules, similar to the clusters of processing units used in HPC systems.

One approach to realize such networks is to use microwave-to-optical quantum transducers~\cite{Fan2018,Higginbotham2018,Forsch2019,Mirhosseini2020}, with which superconducting circuits may be entangled with optical photons to communicate over long distances in a fashion similar to single atoms~\cite{Moehring2007}, trapped ions~\cite{Hofmann2012}, or defects in diamond~\cite{Bernien2013}.
However, despite the constant improvement of microwave-to-optical transducers, bringing their conversion efficiency, bandwidth, added noise, laser-induced quasiparticle poisoning and heat loads to practical levels on a single device remains a challenge.

A complementary approach is to connect dilution-refrigerator based cryogenic systems with cold, superconducting waveguides~\cite{Xiang2017}.
This approach could prove advantageous to distribute quantum computing tasks in local cryogenic quantum networks, as it would benefit from readily available, fast, deterministic, error-correctable and high-fidelity, chip-to-chip quantum communication schemes with microwave photons~\cite{Xiang2017,Kurpiers2018,Campagne-Ibarcq2018,Axline2018,Zhong2019,Leung2019,Chang2020a,Burkhart2020}.
In this article, we report the realization of such a cryogenic quantum microwave channel between superconducting qubits located in two distinct dilution refrigerator units. Using a photon shaping technique to transfer excitations deterministically~\cite{Cirac1997,Kurpiers2018}, we transfer qubit states and generate entanglement on-demand between the distant qubits.

\begin{figure}[t]
\includegraphics[width=1\columnwidth]{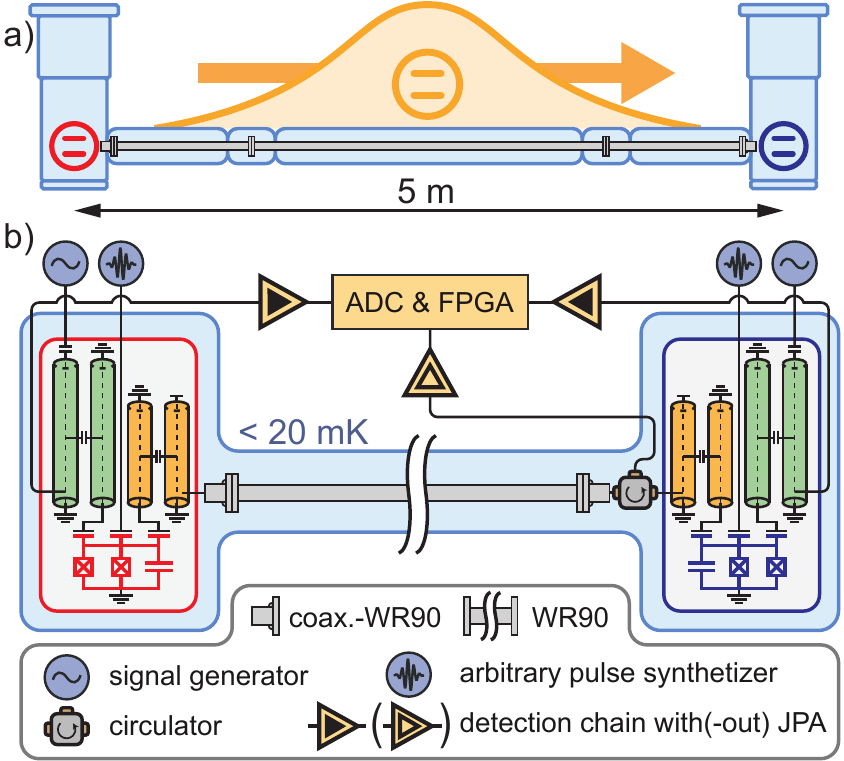}
\caption{
a) Schematic representation, and b), simplified circuit diagram of the experimental setup. Each transmon qubit, at node A (red) and B (blue), is connected to two Purcell filtered $\lambda/4$ resonators: one for readout (green) and one for excitation transfer by emission of a shaped photon (yellow). The light blue background illustrates the refrigerated space.
}
\label{fig1}
\end{figure}

Our experimental setup consists of two cryogen-free, dilution refrigerators, each of which houses a superconducting circuit with a single qubit cooled to below $20\,{\rm {mK}}$ temperature, and  separated by $5\,{\rm {m}}$ (\cref{fig1}). The two identically designed circuits have a frequency-tunable transmon qubit, each with relaxation and coherence times $T_{1}\simeq 12\,\mu{\rm {s}}$ and $T_{2}^{e}\simeq 6\,\mu{\rm {s}}$, coupled dispersively to two Purcell filtered resonators: one for readout, and one for excitation transfer, shown in green and yellow, respectively, in \cref{fig1}~b. The $\ket{g}$ to $\ket{e}$ transition frequencies of transmon qubits labeled A and B are tuned to $\omega_{{\rm q,A}}/2\pi=6.457\,{\rm {GHz}}$ and $\omega_{{\rm q,B}}/2\pi=6.074\,{\rm {GHz}}$, respectively, by applying a magnetic field to their SQUID loops. This adjusts the transfer resonators to the same frequency $\omega_{t}/2\pi=8.406\,{\rm {GHz}}$ through the dispersive shift~\cite{Dickel2018}. Here, $\ket{g}$, $\ket{e}$ and $\ket{f}$ denote the three lowest energy levels of the transmon qubit. 

We connect the transfer resonators to each other through a $4.9\,{\rm {m}}$ long, superconducting, rectangular aluminum WR90 waveguide, in series with two flexible, coaxial copper cables of $0.4\,{\rm {m}}$ length each and a circulator. At millikelvin temperatures, the waveguide exhibits attenuation below $1\,{\rm {dB}/{\rm {km}}}$ over the X band ($8$\textendash $12\,{\rm {GHz}}$), which amounts to a total loss below $10^{-3}$ over $4.9\,{\rm {m}}$ of waveguide (\cref{secWGLoss}).
With attenuation levels comparable to that of optical photons in telecom fibers~\cite{Tamura2017}, the waveguide is in principle suited for high-fidelity transmission of microwave photons over intra-city scale distances~\cite{Xiang2017}.

To perform single-qubit gates, we apply microwave pulses created by arbitrary waveform generators to each qubit through dedicated drive lines. 
To perform readout, we apply a gated microwave tone to the readout resonator. The transmitted signal is then amplified, down-converted, digitized, and processed by a field programmable gate array (FPGA). Using quantum-limited Josephson parametric amplifiers (JPA) in the detection chain, we achieve single-shot three-level discrimination of the transmon states with $\sim5\,\%$ average error ($10\,\%$ for joint two-qubit readout).
Devices, microwave setup, pulse calibration and qubit readout are discussed in more detail in \cref{appChip,appCalib,appSRO}.

\begin{figure}[t]
\includegraphics[width=1\columnwidth]{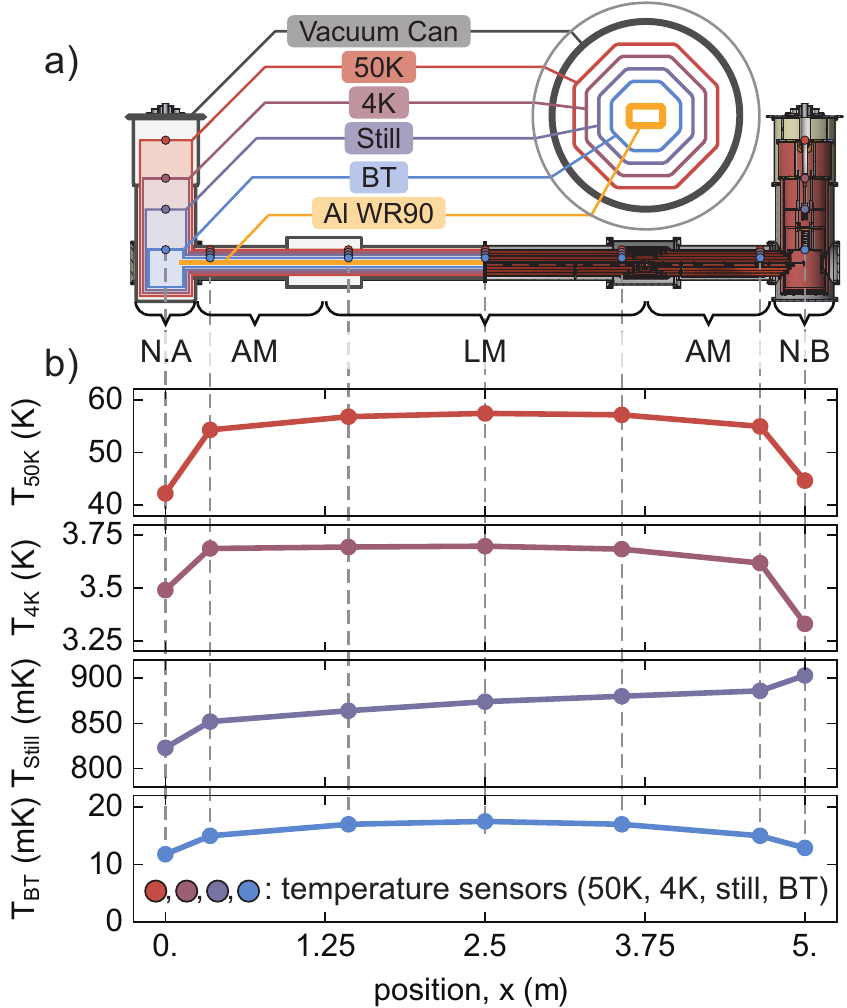}
\caption{
a) Longitudinal cross-section of a schematic representation (left half) and a 3D model (right half) of the cryogenic system. The inset on the top right shows a transverse cross-section of the link. 
b) Measured temperature in steady-state \emph{vs}~sensor position $x$ on the axis along the link for all four temperature stages. Node A/B: N.A/B, adapter module: AM, link module: LM.}
\label{fig2}
\end{figure}

We cool the waveguide to temperatures below $20\,{\rm {mK}}$ by mounting it in a custom-made cryogenic system consisting of concentric, octagonal, radiation shields held at temperatures of approximately $50\,{\rm {K}}$, $4\,{\rm {K}}$, $850\,{\rm {mK}}$ (still), and $15\,{\rm {mK}}$ (base temperature), installed in an o-ring sealed vacuum can (\cref{fig2}~a), see \cref{appCL5Pics} for a photograph of the full system.
The waveguide is thermalized to the base temperature shield every $0.25\,{\rm {m}}$ using flexible copper braids, and the radiation shields are cooled to their equilibrium temperatures using the dilution refrigerators at each end of the system.

The largest heat load on the system is due to room temperature black body radiation, which we mitigate by a set of low-emissivity radiation shields manufactured from high thermal conductivity copper, mechanically supported by thin-walled low thermal conductivity posts. In addition, the heat load on the 50 K stage is reduced by using multi-layer insulation~\cite{Parma2015}.
Generally, minimizing the heat flow between shields at different temperature stages and maximizing the thermal conductivity along each stage reduces thermal gradients and thus allows for lower final temperatures. By making the arrangement of shields light-tight, the base temperature shields cool to below $20\,{\rm {mK}}$. 

We designed the link to be modular, with $1.25\,{\rm {m}}$ long adapter modules to connect the link to each dilution refrigerator and $2.5\,{\rm {m}}$ long link modules, which also allow for an extension of the link (\cref{fig2}~a). To compensate for thermal contraction during cool-down, we use flexible copper braids for thermal anchoring between modules. For the same reason, flexible coaxial cables are used to connect the samples to the waveguide.

To monitor the temperature profile of the link, we installed temperature sensors at the positions indicated in \cref{fig2}~a. Three days after commencing cool-down, the system reaches the steady-state temperature distribution shown in \cref{fig2}~b, demonstrating excellent performance of the system. As expected, on each stage, the temperature is lowest at the nodes and the highest in the middle of the link, with an exception for the still stage where we heated node B to $900\,{\rm {mK}}$ to optimize cooling power by increasing the flow of $^{3}$He.

\begin{figure}[t]
\includegraphics[width=1\columnwidth]{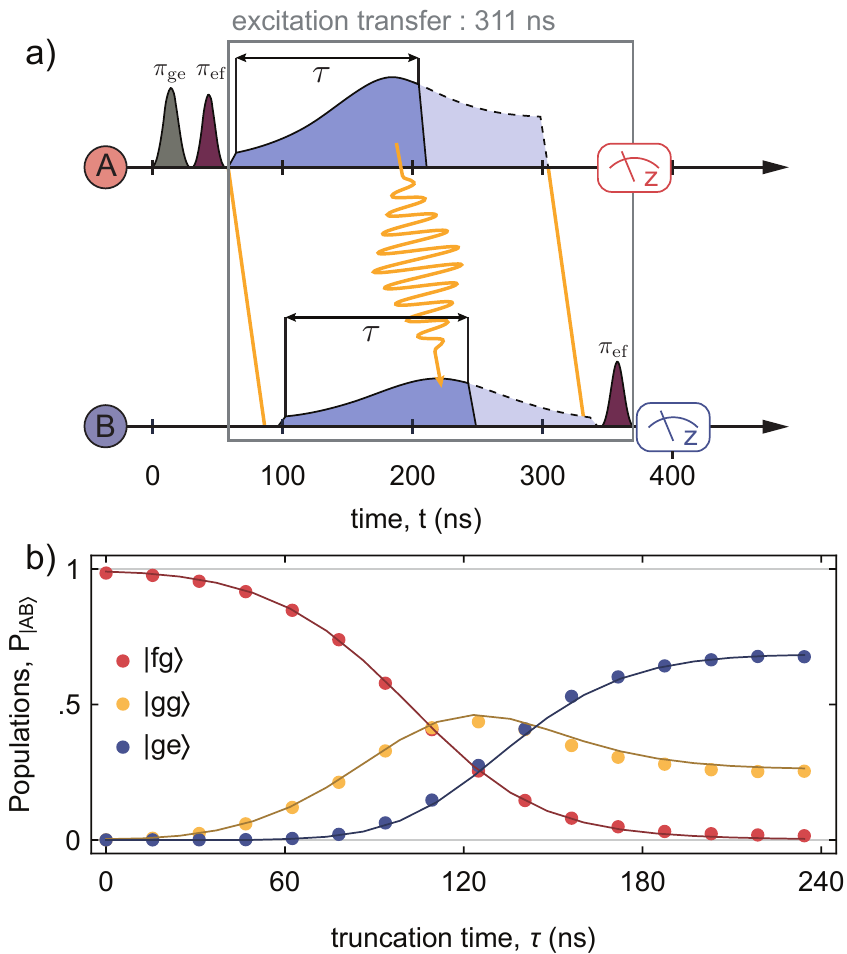}
\caption{ 
a) Pulse scheme used to characterized the excitation transfer dynamics. The $\ket{f0}\rightarrow\ket{g1}$ drives and the $g$-$e$ and $e$-$f$ $\pi$-pulses are represented in blue, grey and bordeaux, respectively. We use solid and dashed lines for the time-truncated ($\tau=140\,{\rm {ns}}$) and the full excitation transfer sequence, respectively. 
The straigth yellow lines illustrate the propagation path of the rising and falling edges of the photon in space-time. 
The sub-sequence defining the excitation transfer is enclosed in a grey box. 
b) Population $P$ in selected two-transmon states $\ket{AB}$ \textit{vs}~$\ket{f0}\leftrightarrow\ket{g1}$ pulse truncation time $\tau$. Solid lines are results of master equation simulations.
}
\label{fig3}
\end{figure}

To characterize the excitation transfer through the link, we first reset the transmon qubits with microwave drives~\cite{Magnard2018} and apply two consecutive gaussian DRAG $\pi$-pulses to prepare the qubit/resonator system at node A in the state $\ket{f0}$ (\cref{fig3}~a), where $\ket{q}$ and $\ket{n}$ in $\ket{qn}$ denote the transmon state and the transfer resonator Fock state, respectively. We then drive transmon A on the $\ket{f0}\leftrightarrow\ket{g1}$ sideband transition~\cite{Zeytinoglu2015,Pechal2014} to populate the transfer resonator with one photon. 
This photon couples into the waveguide at rate $\kappa_{A}/2\pi=8.9\,{\rm {MHz}}$ and propagates to node B in $28\,{\rm {ns}}$, as estimated from the waveguide length and the relevant group velocities (see \cref{appCalib}). We shape the $\ket{f0}\leftrightarrow\ket{g1}$ pulse appropriately to emit the photon with a time-symmetric envelope $\phi(t)\propto\sech(\Gamma t/2)$~\cite{Pechal2014,Kurpiers2018,Morin2019}, where the photon bandwidth $\Gamma$ can be adjusted up to a maximum value of $\min[\kappa_{A},\kappa_{B}]$. 
Here we choose $\Gamma/2\pi=\kappa_{B}/2\pi\simeq6.2\,{\rm {MHz}}$ to minimize the duration of the protocol. 
To absorb the photon at node B, we then drive transmon B with an $\ket{f0}\leftrightarrow\ket{g1}$ pulse whose time-reverse would emit a photon indistinguishable from the incoming one~\cite{Cirac1997}. 
Finally, we apply an $e$-$f$ $\pi$-pulse on transmon B to map the excitation back to the $g$-$e$
manifold, and then perform single-shot read out on both qutrits. Here and in following experiments, we present data which is corrected for readout errors using reference measurements (see \cref{appSRO}).
For these parameters, the excitation transfer sequence, consisting of the $\ket{f0}\leftrightarrow\ket{g1}$ pulses and the final $e$-$f$ $\pi$-pulse, completes in $311\,{\rm {ns}}$.

Truncating the $\ket{f0}\leftrightarrow\ket{g1}$ pulses prematurely at time $\tau$, we characterize the time dependence of the state population of the two transmon qubits throughout the transfer pulse (\cref{fig3}).
As the excitation transfers from node A to node B via the photonic modes (the waveguide and both transfer resonators), the population swaps from the state $\ket{fg}$ of the two spatially separated qubits $\ket{AB}$ to $\ket{ge}$ via the intermediate state $\ket{gg}$.
The final two-transmon state populations highlight the different sources of errors in the excitation transfer. The $\sim3\%$ residual population measured in both $\ket{gf}$ and $\ket{eg}$ (not shown) is due to $e$-$f$ decay. In case of photon loss or failed absorption during the transfer process, the system ends up in the state $\ket{gg}$. 
Comparing the fields of photons emitted from A or B, and detected by the photon measurement chain behind the circulator (\cref{fig1}~b), we determine $22.3\,\%$ photon loss, dominated by the insertion loss of the circulator, and $4.2\,\%$ absorption inefficiency (see \cref{secPhotonLoss}), in reasonable agreement with the $25.3\,\%$ residual population measured in the state $\ket{gg}$. Finally, the transfer efficiency is characterized by the $67.5\,\%$ final population in $\ket{ge}$. 
The time between the applications of the emission and absorption pulses is set to experimentally maximize the transfer efficiency. By comparing the time-of-arrival of photons emitted from A or B in the photon measurement chain, we determine this optimal time difference to be $38\,{\rm {ns}}$, which decomposes into the photon propagation time and an extra $10\,{\rm {ns}}$ lag, as discussed in \cref{appCalib,secPhotonLoss}.
Simulations of the transfer dynamics, using the master equation model from Ref.~\cite{Kurpiers2018} and independently measured parameters, are in good agreement with the data (solid lines in \cref{fig3}~b) and the measured pulse timing (\cref{appCalib}).

\begin{figure}[t!]
\includegraphics[width=1\columnwidth]{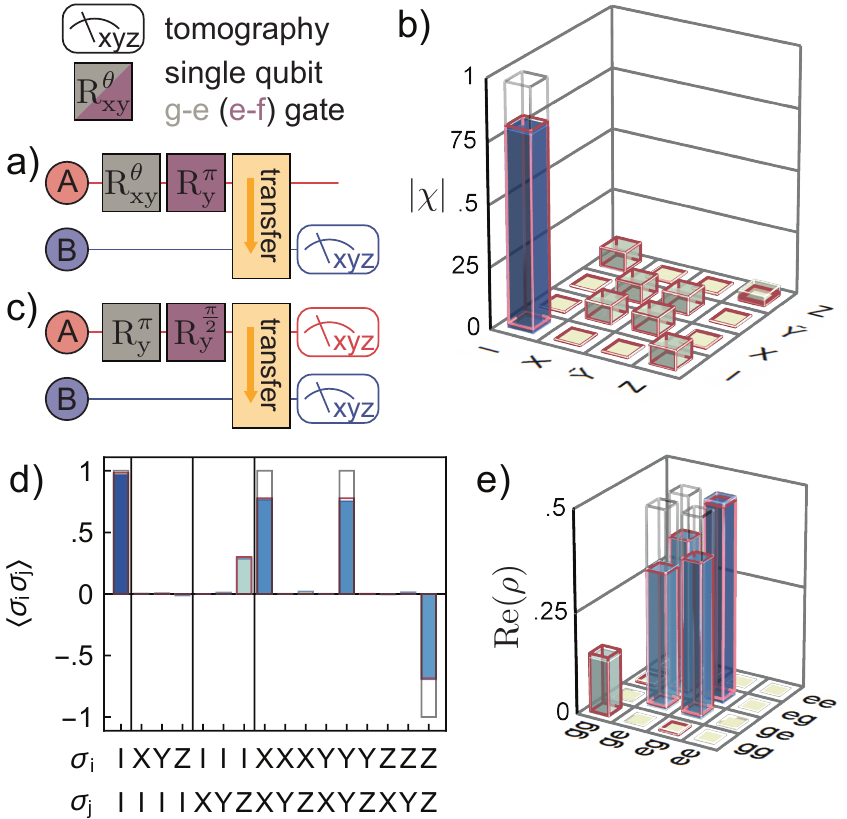}
\caption{ 
a) Quantum circuit used to perform and characterize the qubit state transfer. 
b) Absolute value of the qubit state transfer matrix $|\chi|$, in the Pauli basis.
c) Quantum circuit used to deterministically generate and characterize the Bell state $\ket{\psi^{+}}$. 
e) Expectation value $\av{\sigma_{i}\sigma_{j}}$ of the two-qubit Pauli operators, and e), real part of the density matrix $\rho$ of reconstructed Bell states. 
In b), d) and e), solid blue bars, red wireframes and gray wireframes are the measured, simulated and target quantities, respectively. }
\label{fig4}
\end{figure}

To probe the quantum nature of the excitation transfer, we characterize the qubit state transfer protocol with quantum process tomography. To do so, we reset the qubits to their ground states, prepare A in one of the six mutually unbiased qubit states~\cite{Enk2007}, apply an $e$-$f$ $\pi$-pulse on A, then apply an excitation transfer sequence (\cref{fig4}~a). For each input state we reconstruct the final state of transmon B with three-level quantum state tomography, from which we infer the transfer process matrix $\chi$ (see \cref{appTomo}).
We determine an average transferred state fidelity $\mathcal{F}_{{\rm {s}}}=82.4\pm 0.06\,\%$ and a qubit state transfer fidelity of $\mathcal{F}_{{\rm {p}}}=\Tr(\chi_{{\rm ideal}}\chi)=75.3\pm 0.1\,\%$ relative to the identity process. When correcting for readout errors, these fidelities reach $85.8\pm0.06\,\%$ and $79.5\pm0.1\,\%$, respectively (\cref{fig4}~b).
On average, the input states have equal population in $\ket{g}$, and $\ket{e}$, which are transferred with fidelity $\sim 1$ and $0.675$, respectively, as the former is insensitive to photon loss and decay, and the latter corresponds to the excitation transfer efficiency.
Therefore the state transfer fidelities $\mathcal{F}_{{\rm {s}}}$ and $\mathcal{F}_{{\rm {p}}}$ lie between these two values.

To generate entanglement across the link, we prepare qubit A in $(\ket{e}+\ket{f})/\sqrt{2}$, qubit B in $\ket{g}$, and apply the excitation transfer pulses (\cref{fig4}~c). Using quantum state tomography, we reconstruct the two-qutrit density matrix $\rho_{3\otimes3}$ of qubits A and B (see \cref{appTomo}).
To quantify the entanglement with standard metrics, we consider the density matrix $\rho$, consisting of the two-qubit elements of $\rho_{3\otimes3}$ (\cref{fig4}~d and e). We determine the fidelity $\bra{\psi^{+}}\rho_{m}\ket{\psi^{+}}=79.5\pm0.1\,\%$ ($71.9\pm0.1\,\%$) with respect to the ideal Bell state $\ket{\psi^{+}}=(\ket{ge}+\ket{eg})/\sqrt{2}$, and evaluate a concurrence of $\mathcal{C}(\rho_{{\rm 2\times2}})=0.746\pm0.003$ ($0.588\pm0.002$), with (without) correction for readout errors.

Simulations of the qubit state transfer and entanglement generation sequences are in good agreement with the measurement results, as quantified by the small trace distances $\sqrt{\Tr(|\chi-\chi_{{\rm sim}}|^{2})}=0.09$ and $\sqrt{\Tr(|\rho-\rho_{{\rm sim}}|^{2})}=0.023$ between the reconstructed and simulated quantities. These simulations suggest that photon loss and transmon decay are the dominant sources of errors in these protocols, contributing to $11.8\,\%$ and $\sim6\,\%$ infidelity, respectively. 
In future experiments, the photon loss may be reduced to $5\,\%$ by removing the circulator~\cite{Zhong2019,Leung2019,Chang2020a,Burkhart2020}, by using a printed circuit board (PCB) metallized with a superconductor, and by using low-loss coaxial cables between the device and the waveguide. Simulations of the protocols with $5\,\%$ photon loss and reasonably improved coherence times ($T_{1}\simeq T_{2}^{e}\simeq30\,\mathrm{\mu s}$) and transfer resonator bandwidth ($\kappa/2\pi=12\,{\rm {MHz}}$) indicate that Bell state fidelities and state transfer process fidelities as high as $96\,\%$ may be achievable, which highlights the potential of the protocols for quantum communication tasks between distant cryogenic nodes.

This realization of a meter-scale, milli-Kelvin temperature, microwave-frequency coherent quantum link and its use for quantum state transfer and entanglement generation suggests a number of directions for future research. For example, we plan to experimentally investigate the distribution of quantum information processing tasks between quantum nodes hosting multiple qubits using a coherent cryogenic network, an essential part of a modular quantum computer architecture~\cite{Cirac1999}.
In addition, the modularity of the cyogenic link demonstrated here offers a straightforward path toward extending the physical distance between nodes by adding modules to the link. Due to the small photon loss in the superconducting rectangular waveguide, cryogenic links covering distances of tens or even hundreds of meters could be realized, primarily limited by financial constraints imposed by the thermal requirements. On such length scales one may investigate non-local physics~\cite{Hensen2015,Bierhorst2018} or non-Markovian waveguide QED~\cite{Dinc2019,Calajo2019b} with superconducting quantum devices, and set the grounds for microwave quantum local area networks~\cite{Xiang2017}.

\begin{acknowledgments}
This work is supported by the European Research Council (ERC) through the `Superconducting Quantum Networks' (SuperQuNet) project, by the National Centre of Competence in Research `Quantum Science and Technology' (NCCR QSIT), a research instrument of the Swiss National Science Foundation (SNSF), by ETH Zurich, by NSERC, the Canada First Research Excellence Fund and by the Vanier Canada Graduate Scholarships.
\end{acknowledgments}

\section*{Author contributions}

The experiment was designed and developed by P.M., S.S., P.K and T.W..
The samples were fabricated by J.-C.B., T.W. and M.G.. 
The cryogenic system was designed, characterized and assembled by J.L., P.K., P.M., S.S., F.M. and J.S.. 
The experiments were performed by P.M., P.K. and S.S.. 
The data was analysed and interpreted by P.M., P.K and A.W..
The FPGA firmware and experiment automation was implemented by P.M., K.R, A.A., J.S. and F.M..
The master equation simulations were performed by B.R., P.M. and P.K..
The manuscript was written by P.M. and A.W..
All authors commented on the manuscript. 
The project was led by A.W..

\putbib
\end{bibunit}

\begin{bibunit}[apsrev4-1]

\widetext
\clearpage

\setcounter{equation}{0}
\setcounter{table}{0}
\setcounter{page}{1}
\setcounter{secnumdepth}{3}
\makeatletter
\renewcommand{\theequation}{S\arabic{equation}}
\renewcommand{\thefigure}{\text{S}\arabic{figure}}
\renewcommand{\thetable}{\text{S}\Roman{table}}
\renewcommand{\thesection}{S\arabic{section}}
\renewcommand{\bibnumfmt}[1]{[S#1]}
\renewcommand{\citenumfont}[1]{S#1}
\makeatother

\onecolumngrid

\begin{center}
\textbf{\large
Microwave Quantum Link between Superconducting Circuits\\ 
Housed in Spatially Separated Cryogenic Systems}

\vspace{0.04cm}

\textbf{\large{Supplemental Material}}
\end{center}

\vspace{\columnsep}
\twocolumngrid

\section{Measurement of waveguide loss}
\label{secWGLoss}

\begin{figure}[ht!]
\includegraphics{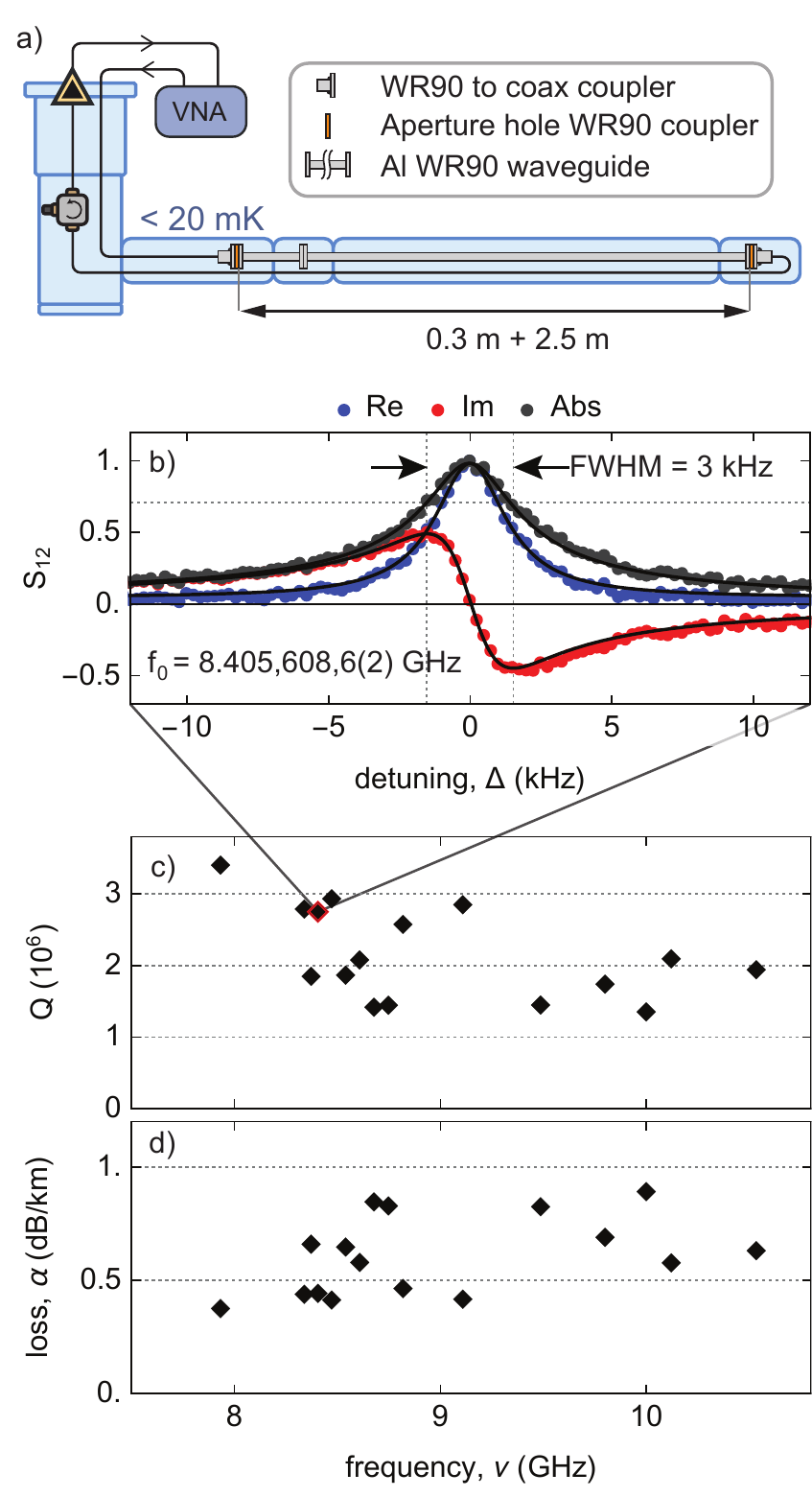}
\caption{
a) Schematic diagram of the waveguide loss characterization experiment, using the same legend as in \cref{fig1}. 
b) Real part (blue), imaginary part (red) and absolute value (grey) of the transmission spectrum $S_{12}$ around a waveguide cavity mode resonance at $f_{0}=8.405,608,6(2)\,{\rm {GHz}}$. 
c) Loaded quality factor $Q$, and d), loss rate $\alpha$ \emph{vs}~resonance frequency $\nu$ for selected waveguide cavity mode resonances.
}
\label{figSWR90}
\end{figure}

To estimate the loss in the rectangular waveguide connecting nodes A and B, we measure the attenuation constant of a similar waveguide using the resonant cavity technique described in Ref.~\cite{Kurpiers2017}. 
The device under test consists of two pieces of rectangular waveguide of the WR90 type, made of AL6061 aluminum without any surface treatment. The two pieces, of $12''$ ($304.8\,{\rm {mm}}$) and $2.5\,{\rm {m}}$, respectively, are joined in a flange-to-flange flat connection and held below $15\,{\rm {mK}}$ inside a dedicated cryogenic system made of a cryogenic node and a $3.75\,{\rm {m}}$ long cryogenic arm (\cref{figSWR90}~a). 
Both ends of the waveguide are closed with aluminum plates with an aperture hole, to form a multi-mode 3D cavity displaying an inter-mode frequency spacing ranging from $30\,{\rm {MHz}}$ to $45\,{\rm {MHz}}$ through the X band. The dimensions of the aperture holes are chosen such that the cavity modes are undercoupled.

Using a vector network analyzer (VNA), we measure selected resonance peaks between $7.5\,{\rm {GHz}}$ and $11\,{\rm {GHz}}$ in transmission, and fit them to a Lorentzian curve to extract their loaded quality factor $Q$ (\cref{figSWR90}~b and c). The loaded quality factor being a lower bound to the internal quality factor $Q_{i}$, we determine an upper-bound to the attenuation constant~\cite{Pozar2012,Kurpiers2017}  
\begin{equation}
\alpha(\nu)=\frac{1}{Q(\nu)}\frac{2\pi\nu}{c\sqrt{1-\left(\frac{c}{2a\nu}\right)^{2}}},\label{eqSAtt}
\end{equation}
where $\nu$ is the resonance frequency, $c$ is the speed of light in vacuum and $a=22.86\,{\rm {mm}}$ is the width of the waveguide.
The attenuation constant is found to be below $1\,{\rm {dB}/{\rm {km}}}$ for all measured data points (\cref{figSWR90}~d). 

\section{Chip fabrication, device parameters and measurement setup}
\label{appChip} 

\begin{figure}[t]
\includegraphics[width=1\columnwidth]{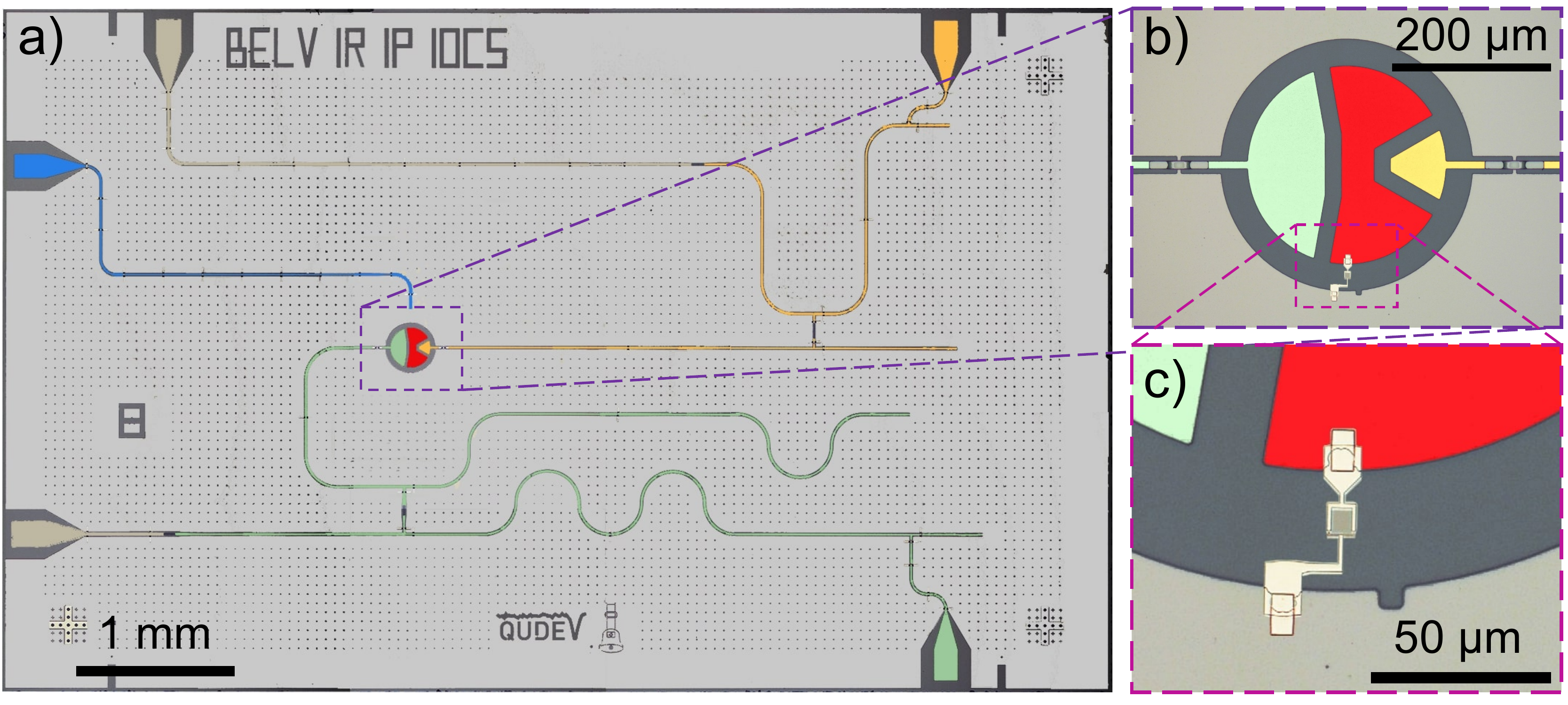}
\caption{
a) False color photograph of a chip similar to those used in the experiment, before deposition of the Josephson junctions, showing the transmon island (red), the drive line (blue), the readout circuitry (green) and the transfer circuitry (yellow). 
b) Microscope image of the transmon qubit, after deposition of the Josephson junctions. 
c) Enlarged view of the Josephson junctions, see scale bars.}
\label{figSSamples}
\end{figure}

We fabricated the two samples on $4.3\,{\rm {mm}\times7\,{\rm {mm}}}$ silicon substrates. 
We patterned the qubit pad and the coplanar waveguide (CPW) structures in a $150\,{\rm {nm}}$ thick niobium film sputtered on a silicon substrate with reactive ion etching in a photolithographic process. 
In a second photolithographic step, we deposited Al/Ti/Al airbridges to connect the ground plane at selected places across the CPWs. We fabricated the Al/AlOx/Al Josephson junctions in a third step with electron-beam lithography and double-angle shadow evaporation. 
Each sample (\cref{figSSamples}) was then mounted, glued and wirebonded to a copper PCB, which was packaged in a copper sample holder. We mounted each sample holder to the base plate of the corresponding dilution refrigerator and wired the devices to the instruments as documented in the wiring diagram \cref{figSCablingDiagram}. 

For each chip, we characterized the parameters of the readout and transfer resonator and Purcell filter circuitry, including resonant frequencies, coupling rates and dispersive shifts, from the transmission spectra through their respective Purcell filtering resonator with the transmon initialized in state $\ket{g}$ or $\ket{e}$, using methods and models similar to those described in Ref.~\cite{Walter2017}. 
Using Ramsey-type experiments, we extracted the transition frequencies and coherence times of the three-level transmon qubits. Both chips have similar parameters (\cref{tabParameterSummary}). 
However, because we tuned qubit A and B to different operating frequencies, qubit frequency dependent parameters such as the qubit/resonator shifts and the $\ket{f0}\leftrightarrow\ket{g1}$ transition frequencies are different at node A and B. The transfer Purcell filter bandwidth differs significantly between the samples, which we suspect to be due to the sensitivity of this parameter to the output impedance.

Using the procedure described in Ref.~\cite{Geerlings2013}, we measured a transmon thermal population at equilibrium of approximately $16\,\%$ for each qubit. We suspect that the absence of infra-red filters in the cables connecting to the samples, and of radiation tight base temperature shields at the node cryostats, leading to poor infra-red shielding, causes this high effective transmon temperature~\cite{Serniak2018}, which we chose to mitigate in future experiments.
We did not determine the residual excitation left after active reset explicitly. However, assuming that the reset is limited by spontaneous rethermalization, we estimate a residual excitation after reset of $0.08\,\%$ ($0.12\,\%$) for transmon A (B) using the analytical expressions from Ref.~\cite{Magnard2018}. 
The fraction of readout errors when the qubit is initialized in $\ket{g}$ gives an upper-bound of $1.3\,\%$ ($0.6\,\%$) to the residual excitation for transmon A (B), see \cref{appSRO}.

\begin{table}[t]
{\footnotesize{}}%
\begin{tabular}{|ll|r|r|}
\hline 
{\footnotesize{}quantity, symbol } & {\footnotesize{}unit } & {\footnotesize{}\hfill{}Node A\hfill{} } & {\footnotesize{}\hfill{}Node B\hfill{}}\tabularnewline
\hline 
\hline 
{\footnotesize{}qubit transition frequency, $\omega_{q}/2\pi$ } & {\footnotesize{}GHz } & {\footnotesize{}6.457 } & {\footnotesize{}6.074 }\tabularnewline
{\footnotesize{}transmon anharmonicity, $\alpha/2\pi$ } & {\footnotesize{}MHz } & {\footnotesize{}-262 } & {\footnotesize{}-262 }\tabularnewline
{\footnotesize{}energy relaxation time on $ge$, $T_{{\rm 1ge}}$ } & {\footnotesize{}$\mathrm{\mu s}$ } & {\footnotesize{}12.2 } & {\footnotesize{}11.7 }\tabularnewline
{\footnotesize{}energy relaxation time on $ef$, $T_{{\rm 1ef}}$ } & {\footnotesize{}$\mathrm{\mu s}$ } & {\footnotesize{}4.9 } & {\footnotesize{}5.0 }\tabularnewline
{\footnotesize{}coherence time on $ge$, $T_{{\rm 2ge}}^{e}$ } & {\footnotesize{}$\mathrm{\mu s}$ } & {\footnotesize{}7.6 } & {\footnotesize{}5.0 }\tabularnewline
{\footnotesize{}coherence time on $ef$, $T_{{\rm 2ef}}^{e}$ } & {\footnotesize{}$\mathrm{\mu s}$ } & {\footnotesize{}7.1 } & {\footnotesize{}5.0 }\tabularnewline
{\footnotesize{}thermal excitation at equilibrium, $n_{{\rm th}}$ } & {\footnotesize{}\% } & {\footnotesize{}16.2 } & {\footnotesize{}16.8 }\tabularnewline
{\footnotesize{}$\ket{f,0}\leftrightarrow\ket{g,1}$ transition frequency,
$\nu_{\mathrm{f0g1}}$ } & {\footnotesize{}GHz } & {\footnotesize{}4.022 } & {\footnotesize{}3.485 }\tabularnewline
\hline 
{\footnotesize{}readout resonator frequency, $\omega_{r}/2\pi$ } & {\footnotesize{}GHz } & {\footnotesize{}4.698 } & {\footnotesize{}4.701 }\tabularnewline
{\footnotesize{}readout Purcell filter frequency, $\omega_{{\rm Pr}}/2\pi$ } & {\footnotesize{}GHz } & {\footnotesize{}4.704 } & {\footnotesize{}4.723 }\tabularnewline
{\footnotesize{}readout resonator/qubit coupling, $g_{r}/2\pi$ } & {\footnotesize{}MHz } & {\footnotesize{}202 } & {\footnotesize{}214 }\tabularnewline
{\footnotesize{}readout circuit dispersive shift, $\chi_{r}/2\pi$ } & {\footnotesize{}MHz } & {\footnotesize{}-4.1 } & {\footnotesize{}-7.9 }\tabularnewline
{\footnotesize{}readout resonator/filter coupling, $J_{r}/2\pi$ } & {\footnotesize{}MHz } & {\footnotesize{}19.9 } & {\footnotesize{}20.0 }\tabularnewline
{\footnotesize{}readout Purcell filter bandwidth, $\kappa_{{\rm Pr}}/2\pi$ } & {\footnotesize{}MHz } & {\footnotesize{}71 } & {\footnotesize{}67 }\tabularnewline
{\footnotesize{}readout resonator eff. bandwidth, $\kappa_{r}/2\pi$ } & {\footnotesize{}MHz } & {\footnotesize{}21.7 } & {\footnotesize{}16.8 }\tabularnewline
\hline 
{\footnotesize{}transfer resonator frequency, $\omega_{t}/2\pi$ } & {\footnotesize{}GHz } & {\footnotesize{}8.406 } & {\footnotesize{}8.406 }\tabularnewline
{\footnotesize{}transfer Purcell filter frequency, $\omega_{{\rm Pt}}/2\pi$ } & {\footnotesize{}GHz } & {\footnotesize{}8.444 } & {\footnotesize{}8.470 }\tabularnewline
{\footnotesize{}transfer resonator/qubit coupling, $g_{t}/2\pi$ } & {\footnotesize{}MHz } & {\footnotesize{}307 } & {\footnotesize{}306 }\tabularnewline
{\footnotesize{}transfer circuit dispersive shift, $\chi_{t}/2\pi$ } & {\footnotesize{}MHz } & {\footnotesize{}-5.75 } & {\footnotesize{}-4.0 }\tabularnewline
{\footnotesize{}transfer resonator/filter coupling, $J_{t}/2\pi$ } & {\footnotesize{}MHz } & {\footnotesize{}20 } & {\footnotesize{}20.8 }\tabularnewline
{\footnotesize{}transfer Purcell filter bandwidth, $\kappa_{{\rm Pt}}/2\pi$ } & {\footnotesize{}MHz } & {\footnotesize{}110 } & {\footnotesize{}155 }\tabularnewline
{\footnotesize{}transfer resonator eff. bandwidth, $\kappa/2\pi$ } & {\footnotesize{}MHz } & {\footnotesize{}8.6 } & {\footnotesize{}6.25 }\tabularnewline
\hline 
\end{tabular}{\footnotesize{}\caption{\label{tabParameterSummary} Device parameters for chips A and B.}
}
\end{table}

\section{Pulse and timing calibration}
\label{appCalib}

\begin{figure}[b]
\includegraphics{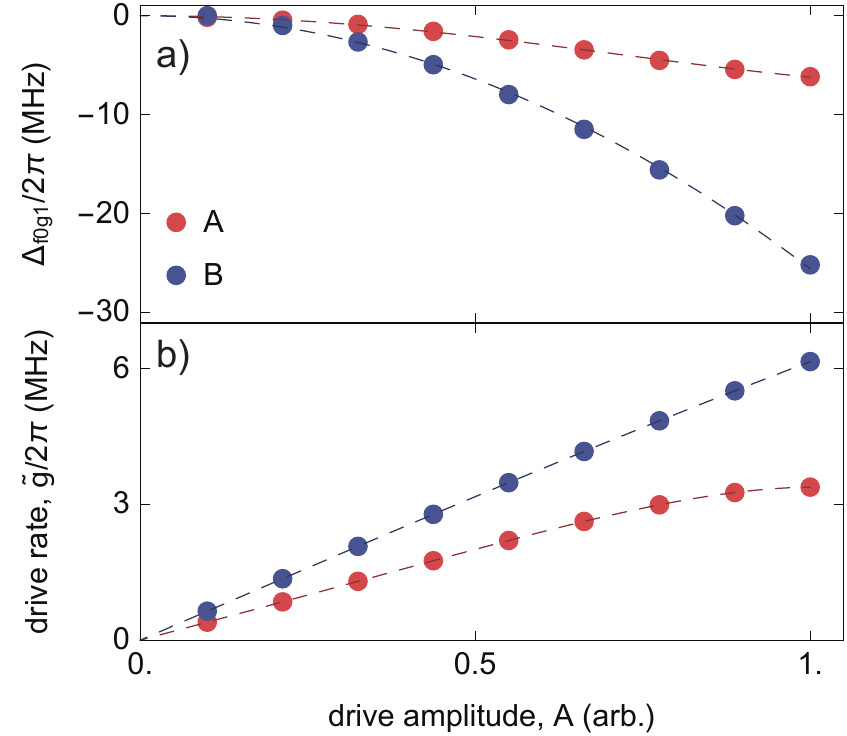}
\caption{
a) ac Stark shift $\Delta_{{\rm {f0g1}}}$, and b) effective drive rate ${\rm {\tilde{g}}}$ \emph{vs}~$\ket{f0}\leftrightarrow\ket{g1}$ drive amplitude $A$. The dashed line in a) and b) are polynomial fits to the data.
}
\label{figSCalib}
\end{figure}

\begin{figure*}[ht!]
\includegraphics{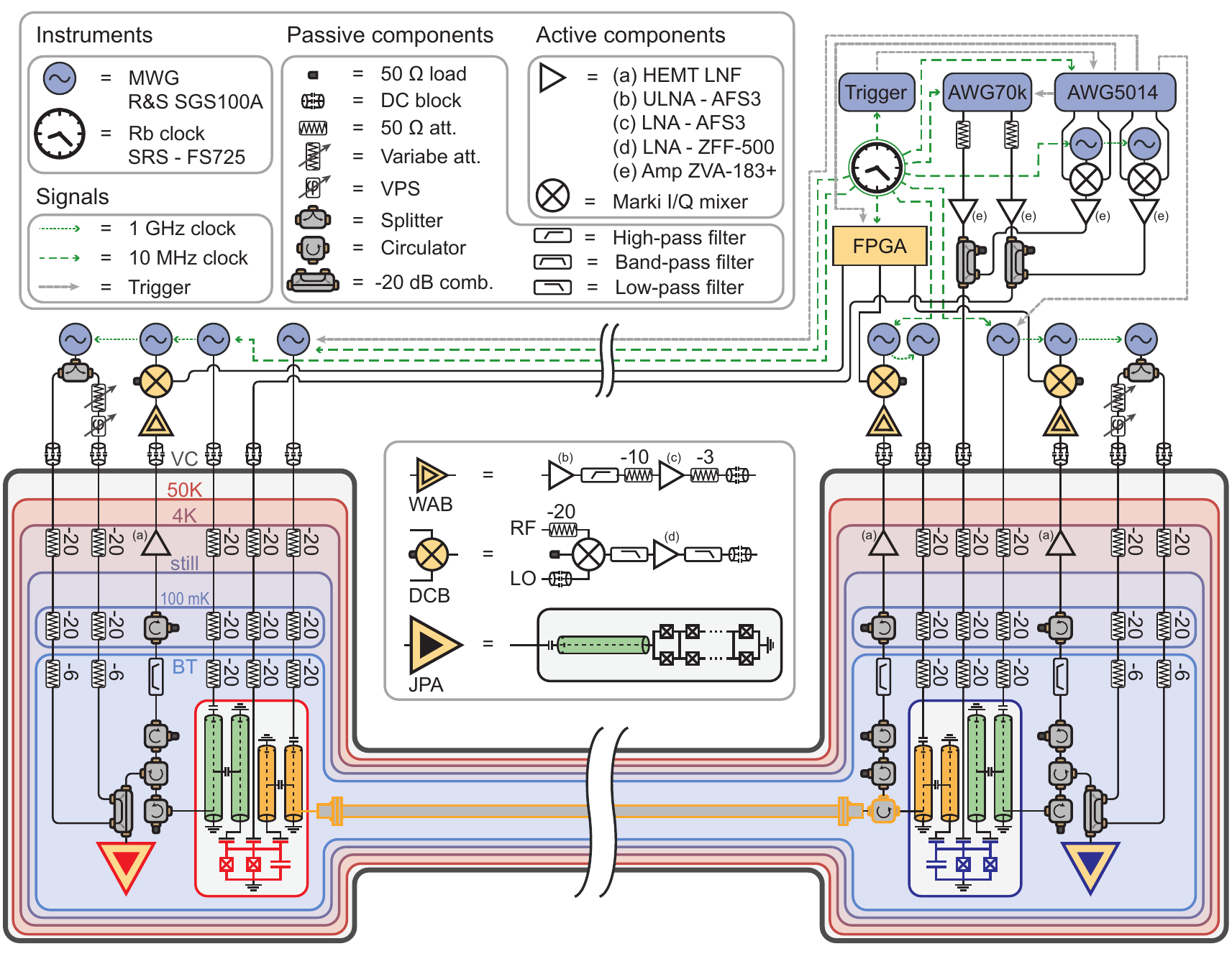}
\caption{
Schematic diagram of the experimental setup. 
WAB: warm amplifier board; DCB: down-conversion board; JPA: Josephson parametric amplifier; MWG: microwave generator; VPS: variable phase shifter.
}
\label{figSCablingDiagram}
\end{figure*}

\begin{figure*}[ht!]
\includegraphics{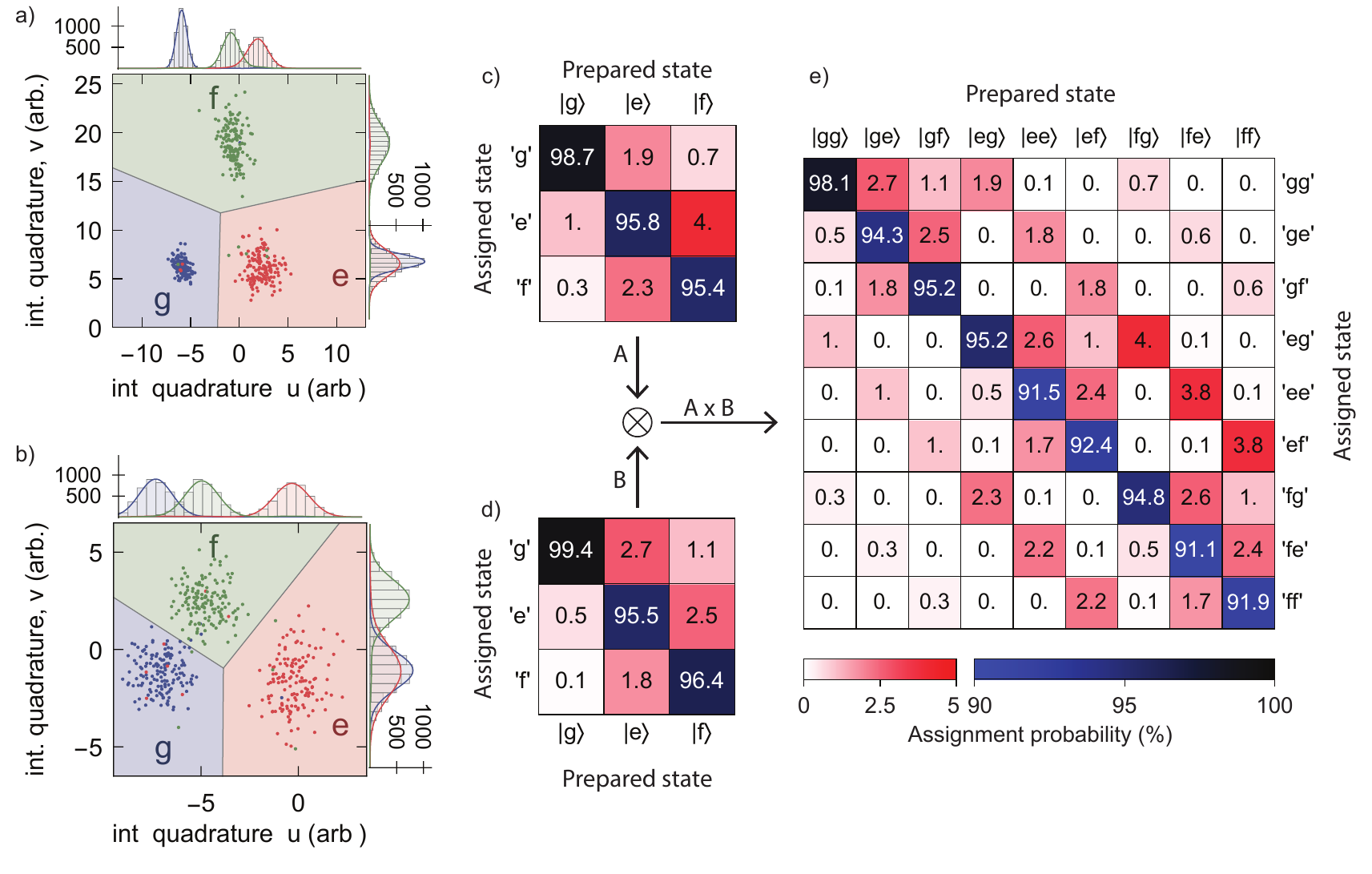}
\caption{
a) and b) Scatter plot of the readout traces, integrated for $248\,{\rm {ns}}$ with optimal weights, with the qubit prepared in state $\ket{g}$ (blue dots), $\ket{e}$ (red dots) and $\ket{f}$ (green dots), for transmon A and B, respectively. The marginal histogram along the integration quadrature $u$ and $v$ is shown for each preparation state on the top and right axes, respectively. Solid-lines are density functions of the marginal three-modal, gaussian distribution estimated from the integrated traces, and scaled to fit the histograms. The $\ket{g}$, $\ket{e}$ and $\ket{f}$ assignment regions are shaded in blue, red and green, respectively. 
c) (resp. d)) Three-state assignment probability matrix extracted from the readout traces and the assignment region displayed in a) (resp. b)), for transmon A (resp. B). 
e) Two-qutrit assignment matrix calculated as the outer product of the single-qutrit assignment matrices. }
\label{figSRO}
\end{figure*}

We use DRAG pulses~\cite{Motzoi2009}, with $28$ and $24\,{\rm {ns}}$ duration, resonant with the $g$-$e$ and $e$-$f$ transitions, respectively, to drive transitions between the three lowest energy states of the transmon qubits.

We apply a microwave tone at frequency $4.249\,{\rm {GHz}}$ ($3.482\,{\rm {GHz}}$) with amplitude $A$ to transmon A (B) to induce an effective drive rate ${\rm {\tilde{g}}}$ between states $\ket{f0}$ and $\ket{g1}$. The $\ket{f0}\leftrightarrow\ket{g1}$ drives are directly synthesized by a separate AWG, then are amplified and combined with the DRAG pulse AWG channel to the drive line of the transmon (\cref{figSCablingDiagram}). 
Using the procedure described in Refs.~\cite{Pechal2014,Magnard2018,Kurpiers2018}, we extract the drive rate ${\rm {\tilde{g}}}$ and the ac-stark shift $\Delta_{f0g1}$~\emph{vs} drive amplitude $A$, and fit each of them with a polynomial function to get a continuous relation between ${\rm {\tilde{g}}}$, $\Delta_{f0g1}$ and $A$ (\cref{figSCalib}). 
This calibration procedure assumes that the transfer resonator decays into a Markovian environment. To realize this condition, we mount a circulator at the far end of the waveguide.

To emit single photons with envelope $\phi(t)\propto\sech(\Gamma t/2)$, we drive the $\ket{f0}\leftrightarrow\ket{g1}$ transition resonantly, with the time-dependent drive rate~\cite{Kurpiers2018,Morin2019}
\begin{equation}
{\rm {\tilde{g}}(t)=\frac{\Gamma}{2}\sech(\Gamma t/2)\frac{1+\frac{1}{2}\left(\frac{\kappa}{\Gamma}-1\right){\rm {e}^{\Gamma t}}}{\sqrt{1+\left(\frac{\kappa}{\Gamma}-1\right)\left({\rm {e}^{\Gamma t}+1}\right)}}.
\label{eqGTilde}}
\end{equation}
If $\Gamma=\kappa$, this choice of photon shape maximizes the achievable photon bandwidth as the rising (falling) edge increases (decreases) with exponential rate $\kappa$, limited by the coupling rate of the absorber (emitter) resonator to the waveguide. 
In this experiment, we minimize the protocol duration by choosing $\Gamma/2\pi=\min(\kappa_{A},\kappa_{B})/2\pi=6.25\,{\rm {MHz}}$. We truncate the $\ket{f0}\leftrightarrow\ket{g1}$ pulse at $\pm4.6/\Gamma$, where the drive amplitude is ramped down to 0 in $6\,{\rm {ns}}$, to emit more than $98\,\%$ of the photon in $246\,{\rm {ns}}$.

For each node, we calibrate the relative time-of-arrival of drive and measurement pulses at the input port of the cryostat, using an oscilloscope. 
Then, to calibrate the relative timing between pulses at node A and pulses at node B, we perform excitation transfer experiments sweeping the time between the application of the emission and absorption pulses. We select the time which maximizes the population transfer to transmon B. 
By comparing the envelopes of photons emitted from node A or B with this timing setting, we infer that the time between the application of the emission and absorption pulses corresponds to the photon propagation time plus a $10\,{\rm {ns}}$ lag (see \cref{secPhotonLoss}). 
This optimal lag is due to the finite truncation of the $\ket{f0}\leftrightarrow \ket{g1}$ pulses. Such a truncated pulse emits a photon whose field amplitude is zero before the pulse start but decays exponentially once the pulse is finished, as the population remaining in the emitting transfer resonator continues to decay into the waveguide. 
Therefore the envelope of the emitted photon is shifted in time compared to the ideal target shape and is better absorbed by a delayed absorption pulse. 
Simulations of the experiment suggest that the excitation transfer is optimized with a $10\,{\rm {ns}}$ lag, in very good agreement with our observations. 
Accounting for the $28\,{\rm {ns}}$ photon propagation delay in the waveguide (as estimated from lengths and group velocities of each section of the waveguide), we deduce that the absorption pulse is applied $38\,{\rm {ns}}$ after the emission pulse.

\begin{figure*}[ht!]
\includegraphics[width=1\textwidth]{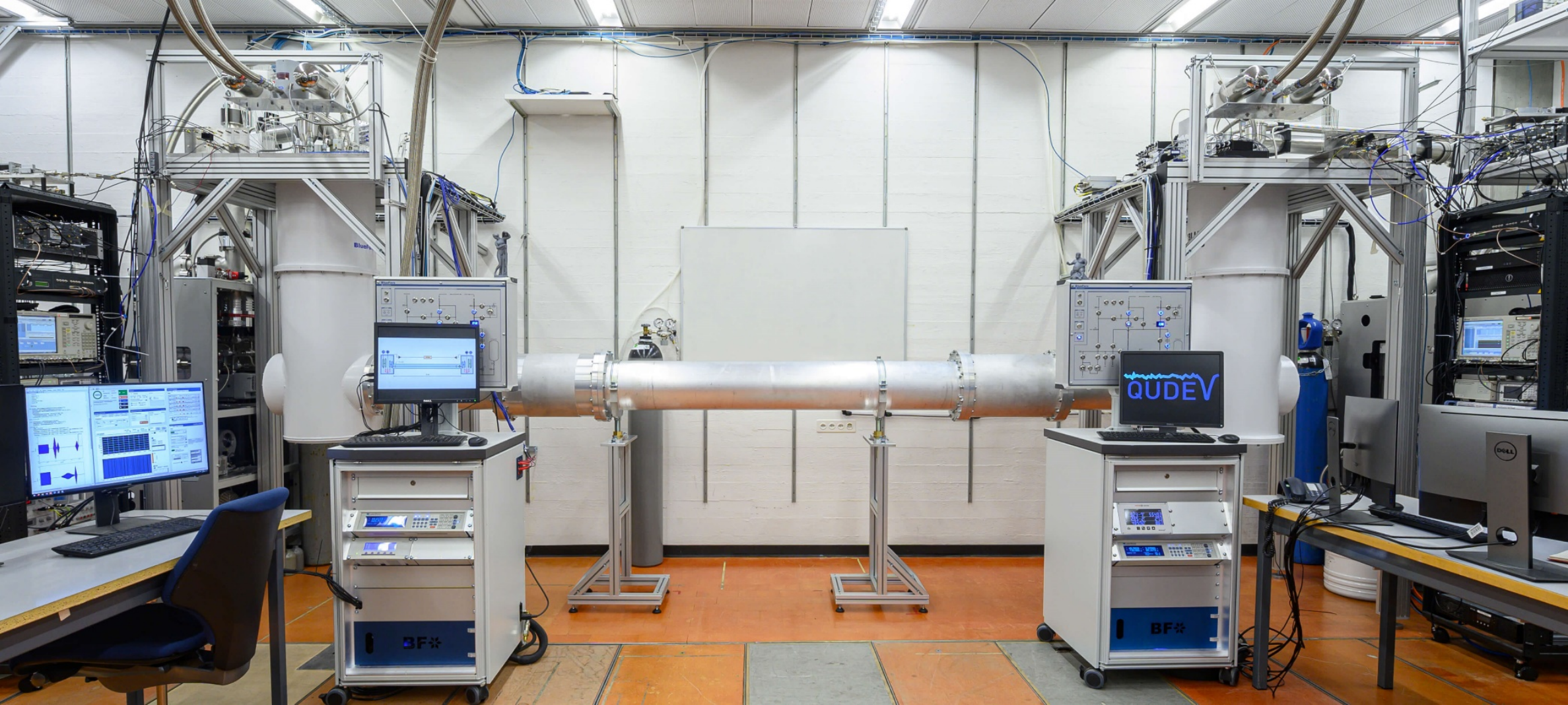}
\caption{
Photograph of the experimental setup during operation. The dilution refrigerator units, whose vacuum cans appear as white vertical cylinders, are connected to each other by the cryogenic link (long horizontal aluminum cylinder). The superconducting qubits and the cold waveguide are housed inside the dilution refrigerator units and the cryogenic link, respectively. The electronic instruments are stored in two black racks seen on the far sides of the picture.
}
\label{figSCL5Pics}
\end{figure*}

\section{Qutrit readout}
\label{appSRO}

To measure transmon A (B), we apply a $4.692\,{\rm {GHz}}$ ($4.680\,{\rm {GHz}}$), gated microwave tone to the input port of the readout Purcell filter. Due to the qubit state dependent dispersive shift $\chi_{r}/2\pi=-4.1\,{\rm {MHz}}$ ($-7.9\,{\rm {MHz}}$) of the readout resonator, the complex amplitude of the transmitted signal carries information about the transmon state, which results in a quantum non-demolition measurement of the transmon~\cite{Walter2017}.

For qutrit state detection, we amplify the signal using a near quantum-limited reflective JPA with $23.6\,{\rm {dB}}$ gain ($21.3\,{\rm {dB}}$) and $14\,{\rm {MHz}}$ bandwidth ($47\,{\rm {MHz}}$), pumped at $4.689\,{\rm {GHz}}$ ($4.668\,{\rm {GHz}}$). We cancel the JPA pump interferometrically at base temperature to avoid pump-induced qubit dephasing and saturation of subsequent amplifiers. We further amplify the signal at the 4K plate with high-electron-mobility transistors (HEMT), then at room temperature with (ultra-)low-noise amplifiers.
The signal is then down-converted to $250\,{\rm {MHz}}$, digitized at $1\,{\rm {Gs}/{\rm {s}}}$, and digitally down-converted to complex DC values by an FPGA using custom firmware (\cref{figSCablingDiagram}).

Over an acquisition window of duration $\tau=248\,{\rm {ns}}$, the FPGA integrates the signal with two sets of weights to reduce the signal to two real-valued components $u$ and $v$. The integration weights are chosen to maximize contrast in the $\{u,v\}$ plane between measurement traces obtained with the qubit initialized in either one of the states $\ket{g}$, $\ket{e}$ or $\ket{f}$. The integrated traces follow a tri-modal gaussian distribution, the parameters of which we estimate with a maximum likelihood approach. Each gaussian mode corresponds to the probability distribution of a measurement trace in the $\{u,v\}$ plane conditioned on the qubit being in a given state during the measurement. The FPGA then assigns the measurement trace to that state with mode center closest in the $\{u,v\}$ plane (see assignment regions in \cref{figSRO}~a and b).

To calibrate the integration weights and mode centers used in the FPGA-based state assignment process, we prepare the transmon in $\ket{g}$, $\ket{e}$ or $\ket{f}$, record 4000 single-shot traces per prepared state, and run the analysis described above on this data set. We determine the single-shot readout assignment probability matrix $R_{A}$ of transmon A (B), shown in \cref{figSRO}~c (d), from the fraction of traces assigned to state $'j'$ when the qubit was prepared in $\ket{i}$. We determine an average readout error probability of $3.4\,\%$ ($2.9\,\%$) from $R_{A}$ ($R_{B}$), and of $6.2\,\%$ from the joint system assignment probability matrix $R_{A}\otimes R_{B}$ (\cref{figSRO}~e). 

We determine the two-transmon state populations at the end of a given pulse sequence from the fraction of single-shot traces assigned to each state. We adjust this population estimate considering the readout errors by multiplying the population vector with the inverse of the two-transmon readout matrix. Because phase drifts of readout instruments lead to an increase in readout assignment error probability, each experiment contains measurements of reference states to estimate the assignment probability matrix at the time of the experiment. We observe an increase of average readout errors to $\sim5\,\%$ and $\sim10\,\%$ for single and two qutrit readout, respectively.

\section{Photograph of the cryogenic system}
\label{appCL5Pics}

A photograph of the cryogenic system used in this experiment is shown in \cref{figSCL5Pics}.

\section{Photon envelope measurements}
\label{secPhotonLoss}

\begin{figure}[ht!]
\includegraphics{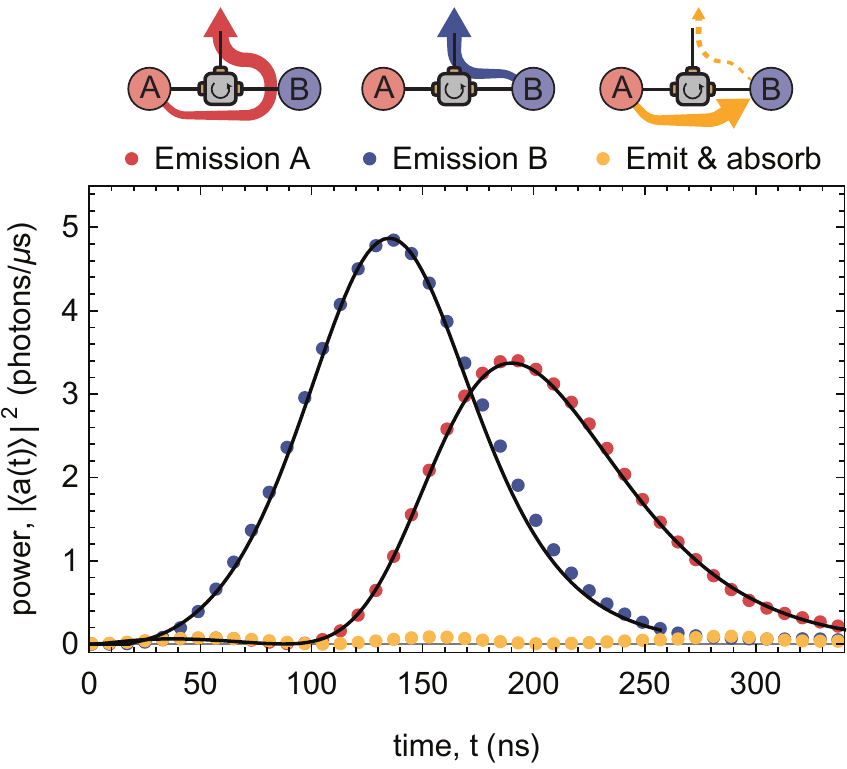}
\caption{
Average electric field amplitude squared $\left|\av{a_{{\rm out}}(t)}\right|^{2}$ \emph{vs}~time of photons emitted from node B (blue), or emitted from node A and reflected from node B in presence (yellow) or absence (red) of an absorption pulse. The $y$-axis is normalized to obtain a unit integrated power for photons emitted from B. The solid lines are results of master equation simulations, with an offset in time to obtain the best agreement with the measurement.
}
\label{figSPhotonMeas}
\end{figure}

To assess the quality of the photon emission and absorption processes, we measure the mean photon field in the photon detection chain after emitting a photon from A, emitting a photon from B, or emitting a photon from A which we absorb at B, as illustrated by the pictograms in \cref{figSPhotonMeas}. In each case, we prepare the emitter qubit in $(\ket{g}+\ket{f})/\sqrt{2}$ and apply an $\ket{f0}\leftrightarrow\ket{g1}$ pulse to emit symmetric-shape photons of state $(\ket{0}+\ket{1})/\sqrt{2}$, with a non-zero average electric field $\av{a_{{\rm out}}}(t)$ proportional to the photon envelope $\phi(t)$.

\begin{figure}[b!]
\begin{minipage}{\textwidth}
\centering
\includegraphics{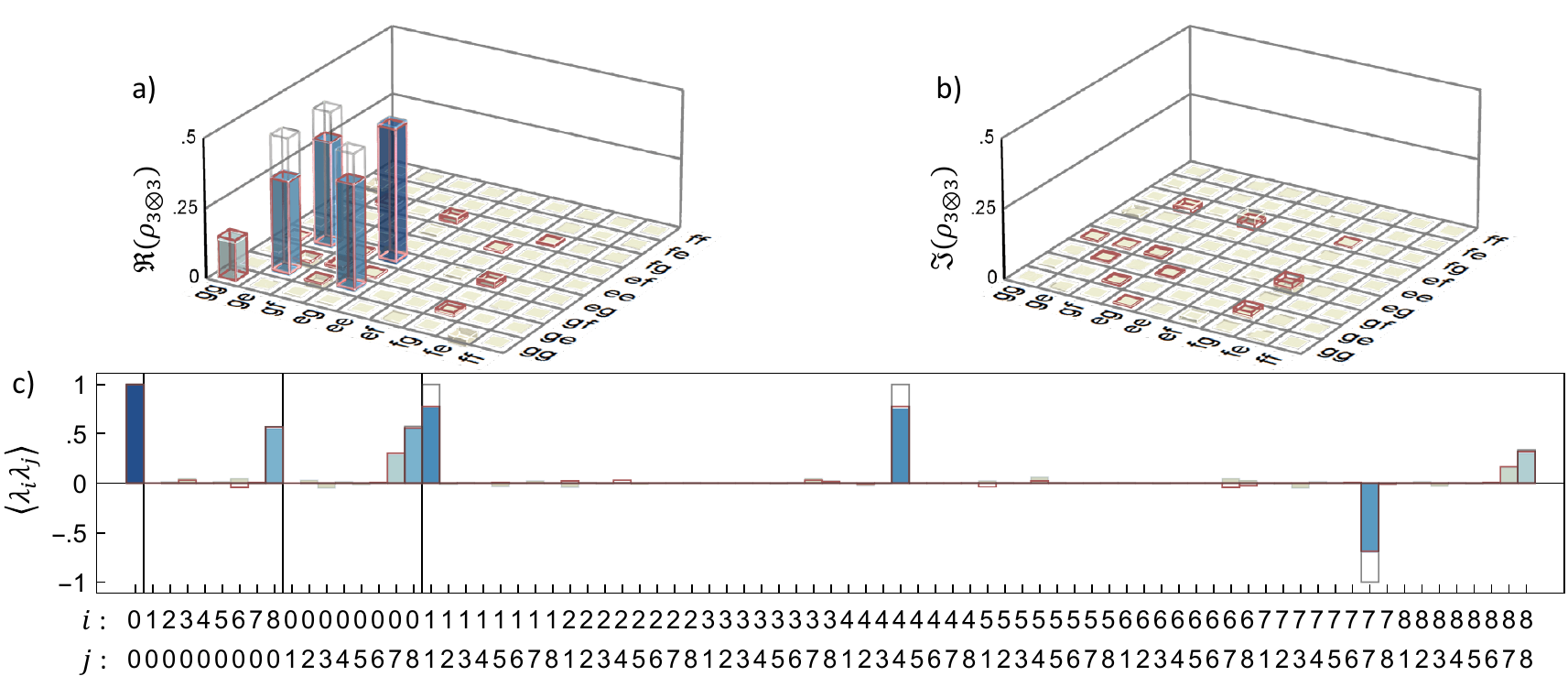}
\caption{
a) Real, and b), imaginary part of the two-qutrit entangled state $\rho_{3\otimes3}$. 
c) Expectation value of the two-qutrit Gell-Mann operators of state $\rho_{3\otimes3}$. 
In all panels, blue solid bars, gray wireframes and red wireframes represent the reconstructed, ideal and simulated values, respectively.
}
\label{figSTomo}
\end{minipage}
\end{figure}

We observe that photons emitted from B have the expected shape and bandwidth, as shown by the close match between the data points and results from master equation simulations using only an offset in time as a fit parameter (solid line in \cref{figSPhotonMeas}). Photons emitted from A  have a shape corresponding to the convolution of a hyperbolic secant shaped envelope with the time-response function of a reflection from resonator B. Here again, simulations agree well with the data.

As discussed in \cref{appCalib}, the offsets in time fitted to the photon emitted from A and B differ by $10\,{\rm {ns}}$. From this we infer that the emission pulse is applied $\Delta \tau_{AB}+10\,{\rm {ns}}$ before the absorption pulse, where $\Delta \tau_{AB}$ is the time it takes the photon to travel from node A to B.

Photons emitted from A have a $22.3\,\%$ lower integrated power $\int\left|\av{a_{{\rm out}}(t)}\right|^{2}dt$ compared to those emitted from B, which corresponds to the probability $l_{AB}$ of losing a photon as it travels from A to B. We use this measured value of $l_{AB}$ in master equation simulations of the experiment.

From the integrated power ratio of the photon field emitted from node A and reflected from node B in presence (yellow), or absence (red) of an absorption pulse at node B, we measure that $95.8\,\%$ of the incoming photon is absorbed by node B.

\section{Tomographic reconstruction}
\label{appTomo}

To perform quantum state tomography of a single qutrit, we measure the three-level population of the transmon with the single-shot readout method described in \cref{appSRO} after applying a tomography gate 
$^{t}{\rm {R}_{{\rm {n}}}^{\theta}}$ selected from the rotation set: $\mathcal{S}=\{
\mathbb{1}$,
$^{{\rm {ge}}}{\rm {R}_{{\rm {x}}}^{\pi/2}}$, 
$^{{\rm {ge}}}{\rm {R}_{{\rm {y}}}^{\pi/2}}$,
$^{{\rm {ge}}}{\rm {R}_{{\rm {x}}}^{\pi}}$, 
$^{{\rm {ef}}}{\rm {R}_{{\rm {x}}}^{\pi/2}}$,
$^{{\rm {ef}}}{\rm {R}_{{\rm {y}}}^{\pi/2}}$, 
$(^{{\rm {ef}}}{\rm {R}_{{\rm {x}}}^{\pi/2}.^{{\rm {ge}}}{\rm {R}_{{\rm {x}}}^{\pi})}}$,
$(^{{\rm {ef}}}{\rm {R}_{{\rm {y}}}^{\pi/2}.^{{\rm {ge}}}{\rm {R}_{{\rm {x}}}^{\pi})}}$,
\clearpage
\noindent$(^{{\rm {ef}}}{\rm {R}_{{\rm {x}}}^{\pi}.^{{\rm {ge}}}{\rm {R}_{{\rm {x}}}^{\pi})\}}}$,
where $t$ denotes the qutrit transition, $\theta$ the rotation angle, and n the rotation axis. We reconstruct the three-level density matrix from this set of measured populations with a maximum likelyhood method, assuming ideal tomography gates. 
This method extends to two-qutrit systems by applying all 81 pairs of local tomography gates from $\mathcal{S}$ on the two transmons, and measuring their population in single-shot.

To characterize the transfer of qubit states from node A to node B, we prepare transmon A in one of the six mutually-unbiased qubit states $\ket{g}$, $\ket{e}$, $(\ket{g}+\ket{e})/\sqrt{2}$, $(\ket{g}+i\ket{e})/\sqrt{2}$, $(\ket{g}-\ket{e})/\sqrt{2}$ or $(\ket{g}-i\ket{e})/\sqrt{2}$~\cite{Enk2007}, and transfer it to transmon B on which we perform full three-level quantum state tomography. Considering only the components of the output density matrices spanned by states $\ket{g}$ and $\ket{e}$, we obtain the two-level process matrix $\chi$ using a maximum likelihood method.

Due to leakage to the $\ket{f}$ level of the transmons during the entanglement protocol, the entangled state cannot be rigorously represented by a two-qubit density matrix. However, to be concise and give standard metrics of entanglement, we reduce the reconstructed two-qutrit state $\rho_{3\otimes3}$ (\cref{figSTomo}) to a two-qubit state $\rho$ consisting of the two-qubit elements of $\rho_{3\otimes3}$. This reduction method leads to states with non-unit trace, but it preserves the fidelity, and gives a conservative estimate to the concurrence compared to a projection of $\rho_{3\otimes3}$ on the set of physical two-qubit density matrices. 

\putbib
\end{bibunit}

\end{document}